\documentclass[12pt]{article}
\begin{document}
\vspace*{.25cm}

\large

\noindent {\Large \bf A quantum-kinetic treatment
for internal \\dynamics of  multilevel atomic systems  moving\\
through a target matter}

\vspace{.5cm}

\hspace{1.4cm} \fbox{A Tarasov} and O Voskresenskaya\footnote{On
leave of absence from Siberian Physical Technical Institute.
Electronic address: voskr@jinr.ru}

\vspace{.5cm}

\hspace{1.5cm}\parbox{13cm}{
{\small Joint
Institute for Nuclear Research, Dubna, Moscow Region, 141980 Russia}
\par\bigskip

{\small {\bf Abstract.} The quantum mechanical consideration of a
passage of relativistic elementary atoms (EA) through a target
matter is given.  A set of quantum-kinetic equations for the density
matrix elements\, describing their internal state evolution at EA
rest frame is derived.\/}

}

\rm
\section{Introduction}

 For the interpretation of the data of DIRAC experiment
 \cite{Dirac04,dirac,nemen} which aims to measure the lifetime of hydrogenlike
EA consisting of $\pi^+$ and $\pi^-$ mesons ($A_{2\pi}$ atoms) one
needs to have the accurate theory for the description of internal
dynamics of the  $A_{2\pi}$ atoms moving through a target matter.

 During their passage through the target $A_{2\pi}$ (pionium atoms)
 interacts with the target atoms that causes the excitation,
 deexcitation or ionization of the $A_{2\pi}$. To describe
 these variations of $A_{2\pi}$ internal states the authors of
 \cite{AT} proposed a set of kinetic equations for the probabilities to
 find the pionium atom in the definite quantum state at some distance
 from the point of $A_{2\pi}$ production.

 It is clear that such ``classical''  description is approximate
 because does not take into account the possible interference (quantum)
 effects. These last can be included in consideration only
 in the framework of a density matrix formalism.

 A kinetic equation for the density matrix of fast
atomic systems passing through a target matter can be given at
target rest frame \cite{voskr,tarvoskr}, but more simple these
equation can be obtained at rest frame of EA \cite{tarvoskr}. A set
of quantum-kinetic equations for the density matrix elements at  EA
rest frame is derived in the present work. A numerically solving
these equations in the first Born approximation is performed in
\cite{santa}.

This paper is devoted to the memory of a remarkable human being and
scientist Alexander Tarasov who  passed away on March 19, 2011.

\section{Derivation of a
quantum kinetic equation for the density matrix }

At the EA rest frame the target moves with the velocity $\vec v_0$,
and the electromagnetic field produced by target atoms is described
by 4-vector potential $A_{\mu}=(\Phi,\vec A)$, $\vec A=(\vec
v_0/c)\Phi$.

The scalar potential $\Phi$ interacts with the charges of mesons and
the vector potential $A_{\mu}$ with their currents.  Because
the typical velocities of the particles forming EA are of
order $\alpha c\ll c$  ($\alpha$ is the fine structure constant),
we will neglect the term proportional to the current in the
Hamiltonian (see \cite{rel}).

Then the internal dynamics of relativistic EA (later, for
definiteness, ``of pionium atoms'') is described by the
Schr\"odinger equation
\begin{equation}
\label{eq:d1} i\frac{\partial\psi(\vec
r,t)}{\partial t}=H\psi(\vec r,t)
\end{equation} with the Hamiltonian of the form
\begin{equation} \label{eq:d2} H=H_{0}+V(\vec r,t),\quad
H_0=T+ V_{0}(\vec r)
\end{equation}
and
\begin{equation}
\label{eq:d3}
T=-\Delta/2\mu=-(d/d\vec r)^2/2\mu\,.
\end{equation}
Here, $V_{0}(\vec r)$  are the potential energy of a pion-pion
interaction and $V(\vec r,t)$ is the potential energy of an
interaction between the pionium and the target atom.

We will suppose that the positions of atoms inside the target are not
varied during the interaction of target with the pionium atom
(the so-called ``frozen'' target approximation).
Then
\begin{equation}
\label{eq:d4}
V(\vec r)=e\sum_i[\Phi\left(\vec r_i(t)-\vec r/2\right)-
\Phi\left(\vec r_i(t)+\vec r/2\right)]\,,
\end{equation}
\begin{equation}
\label{eq:d5}
\vec r_i(t)=\vec r_i(t_0)+\vec v_0(t-t_0)\,,
\end{equation}
\begin{equation}
\label{eq:d6}
\Phi(\vec R)=\gamma\Phi_0\sqrt{\vec
R^2+\gamma^2(\vec v_0\vec R)^2}\,,
\end{equation}
\begin{equation}
\label{eq:d7}
\vec R=\vec r_i(t)\mp\vec r/2,\quad
\gamma=1/\sqrt{1-v_0^2/c^2}\,.
\end{equation}
Here, $\Phi_0$ is the potential of the target atom at it's rest frame,
and we have put the origin of the coordinate system to the
center-of-mass of pionium.

Thus, the solution of the  Schr\"odinger equation  (\ref{eq:d1})
depends on the ``frozen'' positions $\vec r_i(t_0)$ of the target
atoms
$$\psi(\vec r,t)=\psi\Bigl(\vec r,t;\{\vec r_i(t_0)\}\Bigl)\,.$$

The density matrix of pionium is defined as follows:
\begin{eqnarray}
\rho(\vec r,\vec r^{\,\prime};t)&=&\Bigl\langle
\psi\Bigl(\vec r,t;\{\vec r_i(t_0)\}\Bigl)
\cdot \psi(\vec r^{\,\prime},t;\{\vec r_i(t_0)\})
\Bigl\rangle_{\{\vec r_i(t_0)\}}\,,
\label{eq:d7.5}
\end{eqnarray}
where $\langle\rangle_{\{\vec r_i(t_0)\}}$ means the averaging over
all possible positions of target atoms.

Let $t_0$ be the point of time when moving target meet the pionium
atom, and $\psi(\vec r, t_0)$ is the value of pionium wave function
at this time. Then at $t>t_0$
\begin{equation}
\psi\Bigl(\vec r,t;\{\vec r_i(t_0)\}\Bigl)=
\int  G\Bigl(\vec r,\vec r_0;t,t_0;\{\vec r_i(t_0)\}\Bigl)
\psi_i(\vec r_0,t_0)
d\vec r_0\,,
\label{eq:d8}
\end{equation}
where $G$ is the Green function of Eq. (\ref{eq:d1}).

According to \cite{fein}, it can be expressed in terms of the path
integral \begin{equation} G(\vec r,\vec r_0;t,t_0;\{\vec r_i(t_0)\}) =
\int D\vec r(t)\exp(iS)\,,
\label{eq:d9}
\end{equation}
with
\begin{equation}
S = S_0+S_1\,,
\label{eq:d10}
\end{equation}

\begin{equation}
S_0=\int\limits_{t_0}^{t}dt^{\prime}L_0
\bigl(\vec v(t ^{\prime}),\vec r(t ^{\prime})\bigl)\,,\quad
S_1=-\int\limits_{t_0}^{t}dt^{\prime}V
\bigl(\vec r(t ^{\prime}),t ^{\prime}\bigl)\,,
\label{eq:d11}
\end{equation}

\begin{eqnarray}
L_0\bigl(\vec v(t ^{\prime}),\vec r(t ^{\prime})\bigl)&=&
\mu\vec v^{\,2}(t ^{\prime})/2-
V_0\bigl(\vec r(t ^{\prime})\bigl)\,,
\label{eq:d12}
\end{eqnarray}

$$\vec v(t ^{\prime})=d\vec r(t ^{\prime})/dt ^{\prime}\,.$$

It can be shown (see  \cite{VOSKR99}) that
\begin{eqnarray}
S_1&=&-\sum_i\left\{\chi\Bigl(\vec b_i+\vec s(t_i)/2\Bigl)-
\chi\Bigl(\vec b_i-\vec
s(t_i)/2\Bigl)\right\}\\&&~~~~~~~~~~~~\times\vartheta(t-t_i)\,,\nonumber
\label{eq:d15}
\end{eqnarray}
where
\begin{eqnarray}
\chi(\vec b_{\pm})=\frac{e}{v_0}\int\limits_{-\infty}^{\infty}\Phi
\left(\sqrt{\vec b_{\pm}^2+z^2}\right)dz\,,
\label{eq:d17}
\end{eqnarray}
\begin{equation}
\vec b_{\pm}=\vec b_{i}\pm\frac{\vec s(t_i)}{2},\quad
t_i=t_0+\frac{\vec v_0\cdot \vec r_i(t_0)}{v_0^2}\,,
\label{eq:d16}
\end{equation}
\begin{eqnarray}
\vec b_i&=&\vec r_i(t_0)_{\bot}=\vec r_i(t_0)-
\frac{\vec v_0\cdot \vec r_i(t_0)}{v_0^2}
\cdot \vec v_0\,,
\label{eq:d19}
\end{eqnarray}
\begin{eqnarray}
\vec s(t_i)&=&\vec r(t_i)_{\bot}=\vec r(t_i)-
\frac{\vec v_0\cdot \vec r_i(t_0)}{v_0^2}
\cdot \vec v_0\,,
\label{eq:d20}
\end{eqnarray}
the Heavyside step function $\vartheta(t)$ is 0 for $t<0$ and 1 for
$t>0$.

Substituting (\ref{eq:d9})-(\ref{eq:d20}) into (\ref{eq:d7.5}) and
performing the averaging over the ``frozen'' positions of the target
atoms with the help of the prescription of \cite{LP81,AS}, one can
get the following representation for the density matrix:
\begin{eqnarray} \rho(\vec r,\vec r^{\,\prime};t)&=&
\int\widetilde{G}(\vec r,\vec r^{\,\prime};\vec r_0,\vec
r_0^{\,\prime}; t,t_0)\nonumber\\&\times &\psi_i(\vec
r_0,t_0)\psi^{\ast}_i(\vec r_0^{\,\prime},t_0) d\vec r_0d\vec
r_0^{\,\prime}\,, \label{eq:d21}
\end{eqnarray}
with
\begin{equation}
\widetilde{G}(\vec r,\vec r^{\,\prime};\vec r_0,\vec r_0^{\,\prime};t,t_0)
= \int D\vec r(t)D\vec r^{\,\prime}(t)\exp(i\widetilde{S}_0-W)\,,
\label{eq:d22}
\end{equation}
\begin{eqnarray}
\widetilde{S}_0&=&\int\limits_{t_0}^{t}dt^{\prime}\left\{
L_0\bigl(\vec v(t ^{\prime}),\vec r(t ^{\prime})\bigl)
-L_0\bigl(\vec v^{\,\prime}(t^{\prime}),\vec r^{\,\prime}
(t^{\prime})\bigl)\right\}\,,
\label{eq:d23}
\end{eqnarray}
\begin{eqnarray}
W&=&v_0\gamma n_0\int\limits_{t_0}^{t}dt^{\,\prime}
\Omega\bigl(\vec s(t^{\prime}),\vec s^{\,\prime}(t^{\prime})\bigl)\,,
\label{eq:d24}
\end{eqnarray}
\begin{eqnarray}
\Omega\bigl(\vec s(t^{\prime}),\vec s^{\,\prime}(t^{\prime})\bigl)&=&
\int d^2b\left\{1-\exp\left(i
\Phi\bigl(\vec b,\vec s(t^{\prime}),\vec
s^{\,\prime}(t^{\prime})\bigl) \right)\right\}\,,
\label{eq:d25}
\end{eqnarray}
\begin{eqnarray}
\Phi\bigl(\vec b,\vec s(t^{\prime}),\vec
s^{\,\prime}(t^{\prime})\bigl)&=& \chi\bigl(\vec b+\vec
s(t^{\prime})/2\bigl)-\chi\bigl(\vec b-\vec
s(t^{\prime})/2\bigl)\nonumber\\ &-&\chi\bigl(\vec b+\vec
s^{\,\prime}(t^{\prime})/2\bigl)+\bigl(\vec b- \vec
s^{\,\prime}(t^{\prime})/2\bigl)\,.
\label{eq:d26}
\end{eqnarray}
Here, $n_0$ is the number of atoms in the unite volume of target at
it's rest frame,
$\vec s$ and $\vec s^{\,\prime}$ are the transverse parts of the
vectors  $\vec r$ and $\vec r^{\,\prime}$.

From Eqs. (\ref{eq:d21})-(\ref{eq:d24}) it easily derive (see
\cite{fein}) the following equation for the density matrix:
\begin{eqnarray}
i\frac{\partial\rho(\vec r,\vec r^{\,\prime};t)}{\partial t}&=&
H_{0}(\vec r)\rho(\vec r,\vec r^{\,\prime};t)-
H_{0}(\vec r^{\,\prime})\rho(\vec r,\vec r^{\,\prime};t) \nonumber\\
&&-iv_0\gamma n_0 \Omega(\vec s,\vec s^{\,\prime})
\rho(\vec r,\vec r^{\,\prime};t)\,,
\label{eq:d27}
\end{eqnarray}
where the last operator term describes the Coulomb interaction
between EA and the target atoms with account of all multiphoton
exchanges.   Using a generalized optical potential
of the form
$V_{opt}(\vec s,\vec s^{\,\prime}) =k\Omega(\vec s,\vec s^{\,\prime})$,
where $k=-iv_0\gamma n_0$, we can represent this term as
$V_{opt}\rho(t)$.

The form of Eq.~(\ref{eq:d27}) is similar to the form of
Eq. (116) in Ref.~\cite{chang} describing the internal
dynamics of multilevel atoms in laser fields, where the last term
$\Gamma\rho$ describes the contribution of the spontaneous
relaxation.

The equations of motion for the density matrix elements
\begin{equation}
\label{eq:4.41}
\rho_{ik}=\int\psi_i^{\ast}(\vec r)\psi_k(\vec r^{\,\prime})
\rho(\vec r,\vec r^{\,\prime})d\vec rd\vec r^{\,\prime}
\end{equation}
looks like as follows:
\begin{equation}
\label{eq:4.43}
\frac{\partial\rho_{ik}(t)}{\partial t}=i\Delta_{ik}\rho_{ik}(t)
-v_0\gamma n_0\sum_{l,m}\Omega_{ik,lm}\rho_{lm}(t) \,,
\end{equation}
where
$$\Delta_{ik}=\varepsilon_k-\varepsilon_i\,,$$
\begin{equation}
\label{eq:4.44}
\Omega_{ik,lm}=\int\psi_i^{\ast}(\vec r)\psi_l(\vec r)
\psi_k(\vec r^{\,\prime})\psi^{\ast}_m(\vec r^{\,\prime})
\Omega(\vec s,\vec s^{\,\prime})d\vec rd\vec r^{\,\prime}\,,
\end{equation}
the EA wave functions $\psi_{i(k)}$ and the binding energies
$\varepsilon_{i(k)}$ obey the Schr\"o\-din\-ger equation
\begin{equation}
\label{eq:4.42}
H_0\psi_{i(k)}=\varepsilon_{i(k)}\psi_{i(k)}\,.
\end{equation}

Taking into account the lifetime $\tau_{i}$  of the EA,
we can obtain
\begin{eqnarray} \label{eq:4.45.5}
\frac{\partial\rho_{ik}(t)}{\partial t}&=&
\left[i(\varepsilon_{k}-\varepsilon_{i})-\frac{1}{2}
(\Gamma_i+\Gamma_k)\right] \rho_{ik}(t)\nonumber\\&&-v_0\gamma
n_0\sum_{l,m}\Omega_{ik,lm}\rho_{lm}(t)\,,
\end{eqnarray}
where  $\Gamma_{i(k)}=1/\tau_{i(k)}$  is the EA levels width
(for details see \cite{voskr}).

An application of the general formalism discussed here
and
in refs. \cite{voskr,tarvoskr} to the DIRAC experiment is considered
in the paper \cite{santa}.

\section*{Acknowledgments}

The authors\, are\, grateful to Leonid Nemenov and  Leonid Afanasyev
for stimulating interest to the work and useful comments.

\end{document}